\documentclass{svjour3}                     
\smartqed  
\usepackage{graphicx}

\usepackage[top=30truemm,bottom=30truemm,left=25truemm,right=25truemm]{geometry}
\usepackage{amsmath}
\usepackage{amsfonts}
\usepackage{bm}
\usepackage{color}
\usepackage{algorithm}
\usepackage{algorithmic}
\usepackage{multirow}
\graphicspath{{./Figures/}}
\usepackage{enumitem}

\newlength\mylen
\newlist{mycases}{enumerate}{1}
\setlist[mycases,1]{label={Model~\arabic*.}, labelwidth=0\textwidth,leftmargin=20pt,rightmargin=20pt,align=left}

\definecolor{blue}{rgb}{0, 0.4470, 0.7410}
\definecolor{red}{rgb}{0.8500, 0.1250, 0.0480} 
\definecolor{green}{rgb}{0.4660, 0.6740, 0.1880}

\journalname{SN Comput. Sci.}
\begin{document}

\title{Reconstructing three-dimensional bluff body wake from sectional flow fields with convolutional neural networks
}


\author{Mitsuaki Matsuo \and Kai Fukami \\
Taichi Nakamura \and Masaki Morimoto \and Koji Fukagata
}


\institute{   Mitsuaki Matsuo, Kai Fukami, Taichi Nakamura, Masaki Morimoto, Koji Fukagata \at
              Department of Mechanical Engineering, Keio University, Yokohama, 223-8522, Japan \\
              \email{fukagata@mech.keio.ac.jp}
              \and
              Kai Fukami \at
              Department of Mechanical and Aerospace Engineering, University of California, Los Angeles, CA 90095, USA\\
}

\date{Received: date / Accepted: date}

\maketitle

\begin{abstract}
{
The recent development of high-performance computing enables us to generate spatio-temporal high-resolution data of nonlinear dynamical systems and to analyze them for a deeper understanding of their complex nature.
This trend can be found in a wide range of science and engineering, which suggests that detailed investigations on efficient data handling in physical science must be required in the future.
This study considers the use of convolutional neural networks (CNNs) to achieve efficient data storage and estimation of scientific big data derived from nonlinear dynamical systems.
The CNN is used to reconstruct three-dimensional data from a few numbers of two-dimensional sections in a computationally
friendly manner. 
The present model is a combination of two- and three-dimensional CNNs, which allows users to save only some of the two-dimensional sections to reconstruct the volumetric data.
As examples, we consider a flow around a square cylinder at the diameter-based Reynolds number $Re_D = 300$.
We demonstrate that volumetric fluid flow data can be reconstructed with the present method from as few as five sections.
Furthermore, we propose a combination of the present CNN-based reconstruction with an adaptive sampling-based super-resolution analysis to augment the data compressibility.
Our report can serve as a bridge toward practical data handling for not only fluid mechanics but also a broad range of physical sciences.
}
\end{abstract}
\section{Introduction}
\label{sec:intro}
Scientific big data has been accumulated owing to the improvement of the computational power following Moore’s law~\cite{Moore1965}.
These trends in high-performance computing suggest that detailed investigations on big data handling can be helpful for a wide range of scientific communities in the future from the viewpoint of efficient storage and exchange of data.
This paper examines a supervised machine-learning-based method to achieve efficient data handling of scientific big data derived from nonlinear dynamical systems with an example of fluid flows.

Fluid flow data can be regarded as one of the examples of big data derived from complex physics with strong nonlinearities and multi-scale nature.
Due to such natures, large-scale numerical simulations are often required for detailed analyses.
For example, let us consider the use of direct numerical simulation (DNS) over an arbitrary three-dimensional domain discretized with 100 points in each direction --- the sum of data points reaches 4 million dimensions since there are three velocity components (${\bm u}=\{u,v,w\}$) and pressure for all computational points of $100^3$.
We hope you can now feel its unbelievable data size with this simple example.
In fact, the number of grid points contained in the data used in this study is much greater than in this example.
Besides, the requirement for the number of computational grid points for DNS is usually in proportion to $Re^{9/4}$, where $Re$ is the Reynolds number, whereby we encounter challenges due to the limitation of computational power~\cite{kajishima2017computational}.

To handle the high-dimensional complex fluid flow data in an efficient manner, machine learning (ML) has been gaining attention~\cite{BNK2020,Kutz2017}.
The ML is good at handling them by accounting nonlinearities into its procedure~\cite{BHT2020}.
For instance, autoencoder-based reduced-order modeling (ROM)~\cite{HS2006} benefits from nonlinear ML~\cite{THBSDBDY2020,novoa2022real} in fluid mechanics~\cite{Milano2002}.
The autoencoder (AE) has a bottleneck structure and is trained by setting the same data for both input and output.
If the model can output the same data as the input, the high-dimensional original input (or output) is successfully compressed into the bottleneck space referred to as the latent space.
This idea has been widely accepted in fluid dynamics~\cite{GKS2020,MFZNF2021,eivazi2022towards,clainche2022improving,yousif2022physics,doan2021auto}.
Omata and Shirayama~\cite{omata2019} applied a convolutional neural network (CNN)-based AE for a wake of NACA0012 airfoil and investigated the temporal behavior of low-dimensional variables.
Hasegawa et al.~\cite{HFMF2020a,HFMF2019} proposed an AE-assisted surrogate model for high-fidelity fluid flow simulations of unsteady laminar wakes around bluff bodies by combining with a long short-term memory (LSTM).
In their ROM formulation, flow field data are first low-dimensionalized using AE and then the LSTM is employed to predict the temporal evolution of the low-dimensionalized latent variables.
Because the decoder part of AE is able to recover the dimension from the latent variables to physical spaces, the temporal evolution of high-dimensional flow fields can be obtained by following that in the low-dimensional space.
This idea has been demonstrated with hyperbolic shallow water equations~\cite{maulik2020reduced}, a cylinder wake at different Reynolds numbers~\cite{HFMF2020b}, and minimal turbulent channel flow~\cite{nakamura2020extension}.
Instead of LSTM, sparse identification of nonlinear dynamics can also be a candidate as a temporal predictor of low-dimensional vector~\cite{fukami2020sparse}.
In addition to the surrogate modeling efforts, there are several studies that focus on the interpretability of latent variables extracted by AE.
Lusch et al.~\cite{lusch2018deep} proposed a customized AE to guarantee a Koopman operator-like form in a latent space to perform coordinate transformation through neural network operations.
A visualization of AE modes on physical space was also attempted with a mode-decomposing CNN-AE~\cite{MFF2019}.
Since it has recently been reported that several machine learning functions such as multi-scale filters~\cite{xu2020multi,LPBK2020}, skip connection~\cite{EMM2019} and transfer learning~\cite{FNF2020} can enhance the compressibility of AE for fluid flow data, applications with more practical situations, e.g., turbulence, can also be expected in near future.

Machine learning has also exhibited its great potential in state estimation of fluid flows~\cite{BEF2019}.
For instance, a multi-layer perceptron-based model has been applied to reconstruct a global flow field from local sensor measurements~\cite{erichson2019,FFT2020}.
These studies have shown that neural networks can reasonably reconstruct flows with the aid of nonlinear activation functions and outperform conventional linear methods, such as Gappy proper orthogonal decomposition (POD)~\cite{ES1995} and linear stochastic estimation~\cite{NFF2021}.
Instead of considering whole field information, low-dimensional representations such as POD coefficients can be a target variable to reconstruct a global flow field in a reasonable manner, as well summarized in Nair and Goza~\cite{NG2020}.
Moreover, the use of mathematical projection with single ML model enables us to deal with local and sparse sensor measurements that can be in motion~\cite{FukamiVoronoi}.
Furthermore, super-resolution analysis, which reconstructs high-resolution data from its low-resolution data, was performed for turbulent flows by Fukami et al.~\cite{FFT2019a,FFT2019b}.
The extension of super-resolution analysis has been ongoing for not only numerical~\cite{maulik2017resolution,LTHL2020,kim2021unsupervised,FFT2020b,yousif2021high,yousif2022super,yousif2022transformer} but also experimental studies~\cite{DHLK2019,MFF2020,CZXG2019}.
These successes of a global flow field estimation from local sensor measurements or low-resolution data indicate that we solely need to keep these input data and ML models to reproduce the high-dimensional original data.
However, there are few studies for three-dimensional fluid data estimation or low-dimensionalization despite the strong demands for a detailed investigation of efficient data handling.

Several machine-learning studies that address three-dimensional reconstruction can be found in other fields~\cite{yousif2022deep}.
Huang et al.~\cite{HLC2019} applied a hybrid CNN-LSTM model to reconstruct the evolution of a three-dimensional fire flame structure from its two-dimensional projections.
They chose a CNN architecture to capture the features of the fire flame because the physical nature of the evolution of flame structures has much in common with the basis of image processing.
Kench and Cooper~\cite{KC2021} proposed the slice GAN (generative adversarial network) to faithfully synthesize three-dimensional $n$-phase media from two-dimensional material micrographs.
It is reported that convolutional layers can efficiently capture complex features of two-dimensional microstructures while preserving geometric information.
In this manner, convolution-based architectures play an important role in capturing complex physical features in space and reconstructing them in three-dimensional space.
However, the fluid flow dataset, which is a representative example of high-dimensional nonlinear systems, 
needs to be investigated in greater detail.

Motivated above, we consider an efficient data handling method for fluid flow examples.
More concretely, we capitalize on a convolutional neural network (CNN) to reconstruct three-dimensional spatial discretized data from a few numbers of two-dimensional sections.
As an example of three-dimensional data, a flow around a square cylinder at the Reynolds number $Re_D=300$ is considered.
In addition, we also propose an adaptive sampling-based super-resolution analysis to achieve more efficient data handling 
for complex fluid flows.
This adaptive sampling enables us to save low-resolution flow data in a computationally efficient manner by focusing on an important portion of a fluid flow field.
The present paper is organized as follows: the schemes of machine learning and proposed adaptive sampling are summarized in section~\ref{sec:method}.
Results and discussion are offered in section~\ref{sec:result}.
We finally provide concluding remarks in section~\ref{sec:conclusion}.

\section{Methods}
\label{sec:method}

\subsection{Dataset}
\label{sec:dataset}

We consider a flow around a square cylinder at Reynolds number ${Re}_D=300$.
The present example contains the complex three-dimensional structure caused by the merge of both two- and three-dimensional separated shear layers~\cite{BA2017}.
The dataset is prepared by direct numerical simulation (DNS).
The incompressible Navier--Stokes equations with a penalization term are numerically solved~\cite{volumePenal1994},
\begin{flalign}
        {\bm{\nabla}} \cdot {\bm u}=0,~~~{\partial_t {\bm u}} + {\bm{\nabla}} \cdot \left({\bm{u}}{\bm{u}}\right)=-{\bm{\nabla}} p + {{ Re}^{-1}_D}{\bm{\nabla}}^2\bm{u}+\lambda \chi\left({\bm u}_b-{\bm u}\right),
\end{flalign}
where ${\bm u}=\{u,v,w\}$ and $p$ are the velocity vector and pressure, respectively, nondimensionalized by the fluid density $\rho$, the length of the square cylinder $D$, and the uniform velocity $U_\infty$.
The penalization term expresses the body with a penalty parameter $\lambda$, a mask value $\chi$, and the velocity of the object ${\bm u}_b$ which is zero in the present case.
The mask value $\chi$ is $0$ outside a body and $1$ inside a body.
The computational domain is set to $\left(L_x \times L_y \times L_z\right)=\left(20D\times20D\times4D\right)$ and the computational time step is $\Delta t_{\rm DNS}=2.5\times10^{-3}$.
A uniform velocity is imposed at the inflow boundary ($x=0$), the convective boundary condition is adopted at the outflow boundary ($x=L_x$), the slip boundary condition is applied at $y=-L_y/2$ and $y=L_y/2$, and the periodic boundary condition is assumed at $z=-L_z/2$ and $z=L_z/2$.
The center of the square cylinder is located $5.5D$ from the inflow boundary.

In this study, our focus is the part of computational volume around the square cylinder such that $\left(12.8D\times4D\times4D\right)$ with the grid numbers of $(N_x^\sharp\times N_y^\sharp\times N_z^\sharp)=(256\times128\times160)$.
For the present investigation, we prepare the snapshots from the DNS with the time interval $\Delta t_{\rm ML} = 0.25$, which corresponds to 100 folds compared to the time step of the simulation, such that $\Delta t_{\rm ML}=100\Delta t_{\rm DNS}$.
We use 1000 snapshots between $0 \leq t \leq 249.75$ for training the baseline model. 
The dependence of the reconstruction performance on the number of training snapshots is also investigated in this study.
For all cases, we use $70 \%$ of 1000 snapshots for training the models and the remaining $30 \%$ for validation.
A hundred test snapshots between $724.75 \leq t \leq 749.75$ are used to assess the machine-learning model.
Hence, the test data is extracted from sufficiently distanced snapshots from the training data.

\subsection{2D-3D convolutional neural networks}
\label{sec:AEscheme}

We use a convolutional neural network (CNN)~\cite{LBBH1998} to reconstruct the three-dimensional flow data.
The CNN excels at extracting features from a large amount of data thanks to its filter operator, and it has been utilized in image processing and classification tasks~\cite{LBH2015}.
In fluid dynamics, the use of CNN is also spreading because the filter sharing allows us to handle the high-dimensional fluid data efficiently~\cite{FNKF2019,champion2019data,KL2020,Fukamipump2022}.

\begin{figure}[t]
	\vspace{0mm}
	\begin{center}
		\includegraphics[width=0.9\textwidth]{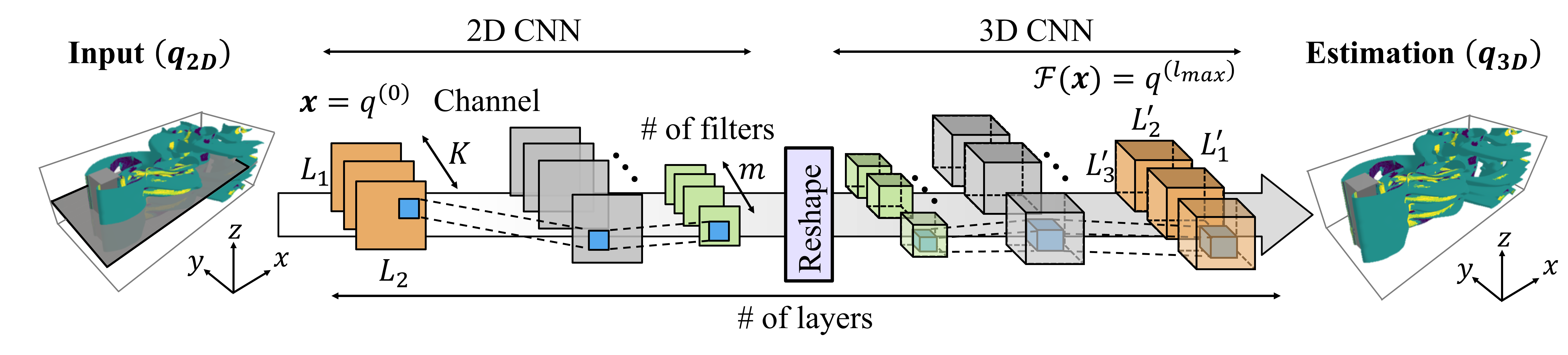}
		\caption{{2D-3D CNN developed in this study. 
            Input cross-sectional data is converted to three-dimensional volumetric data via two-dimensional and three-dimensional convolutional neural networks.}}
		\label{fig:schematic_of_model}
	\end{center}
\end{figure}

A combination of two- and three-dimensional CNN is utilized to reconstruct a three-dimensional flow field from two-dimensional inputs, as shown in figure~\ref{fig:schematic_of_model}.
The two-dimensional CNN is composed of two-dimensional convolutional layers and max pooling layers.
The convolutional layer is used to extract the feature of two-dimensional sections while reducing the dimension of the data with the max pooling layers.
The use of a dimensional reduction tool for this part is vital for the CNN-based data reconstruction to obtain robustness against the rotation and translation of the images and noisy inputs~\cite{HZRS2016}, although the model eventually aims to reconstruct the higher dimension data.
As shown in figure~\ref{fig:schematic_of_model}, the three-dimensional layers are in charge of dimension extension.
Compared to autoencoder-based fluid flow data compression with non-interpretable latent vector~\cite{glaws2020deep,momenifar2022dimension}, the present 2D-3D CNN can store three-dimensional data as the interpretable two-dimensional form such as two-dimensional cross-sections.

\begin{figure}
    \vspace{0mm}
	\begin{center}
		\includegraphics[width=0.94\textwidth]{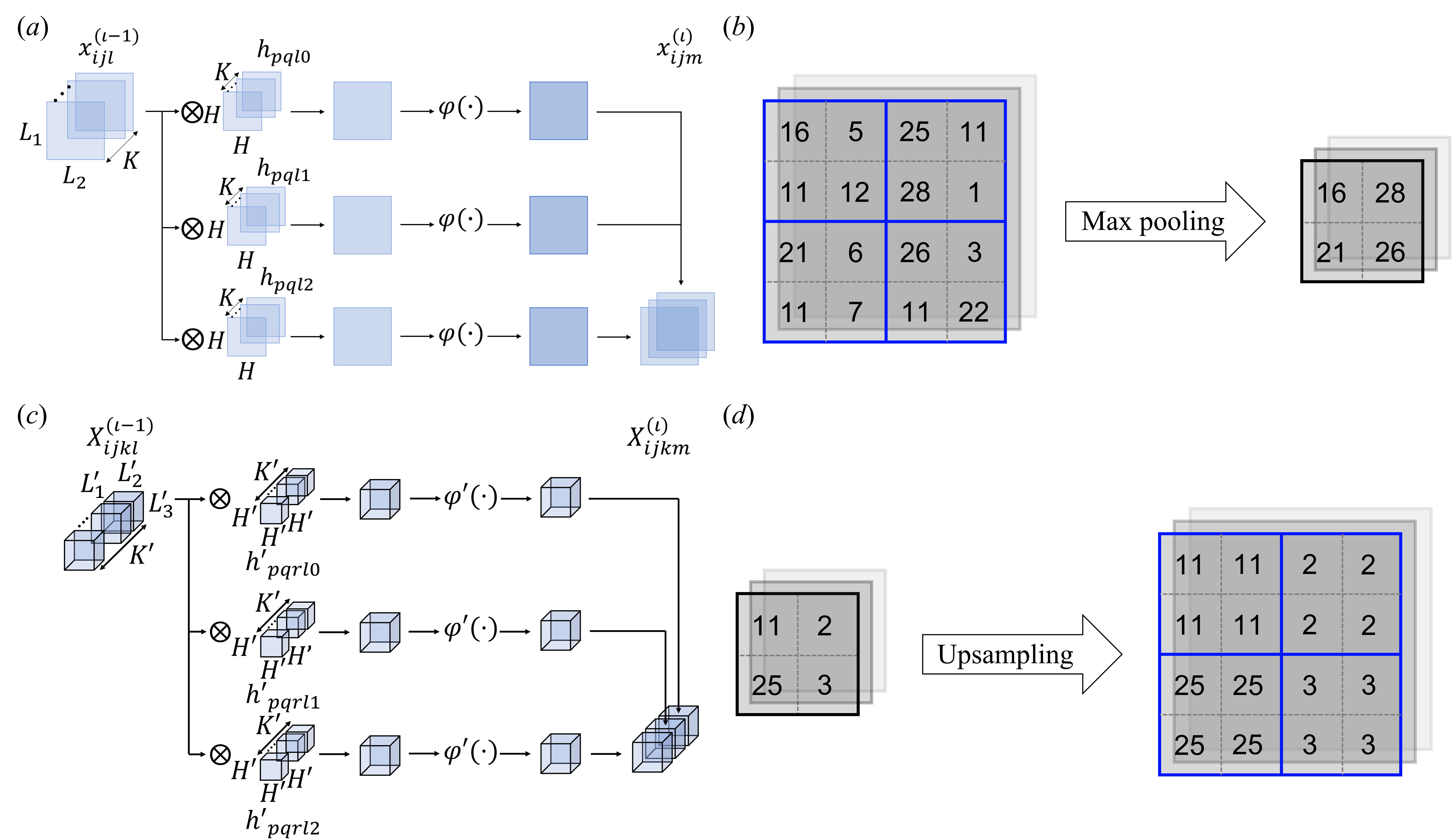}
		\caption{Internal procedure of convolutional neural network. $(a)$ Two-dimensional convolution operation. $(b)$ Max pooling. $(c)$ Three-dimensional convolution operation. $(d)$ Upsampling. We use two-dimensional examples for $(c)$ and $(d)$ for clarity of illustration.}
		\label{fig:CNN_ope}
	\end{center}
\end{figure}

The internal procedure of the two-dimensional convolutional layer is illustrated in figure~\ref{fig:CNN_ope}$(a)$.
In the diagram, the input $x^{(\iota-1)}$ of size $L_1 \times L_2$ at the $(\iota-1)$th layer has $K$ channels corresponding to variables per each position of data.
The filter $h^{(\iota)}$ of size $H \times H$ with $L$ channels is shared for the whole image field of the input.
Through the convolutional operation, the output $x^{(\iota)}$ at the $\iota$th layer, pixel indices $(i,j)$ on a two-dimensional image, and filter index $m$, can be obtained as
\begin{equation}
    x^{(\iota)}_{ijm} = {\varphi}\biggl({b_m^{(\iota)}}+\sum^{L-1}_{l=0}\sum^{{H}-1}_{p=0}\sum^{{H}-1}_{q=0}h_{p{q}lm}^{(\iota)} x_{i+p{{-G}},j+{q{-G}},l}^{(\iota-1)}\biggr),
    \label{eq:2DCNN}
\end{equation}
where $G={\rm floor}(H/2)$, $b_m$ is the bias, and $\varphi$ is an activation function. 
As the activation function $\varphi$, we use a ReLU function~\cite{NH2010} to avoid vanishing gradients.
We also consider a max pooling layer shown in figure~\ref{fig:CNN_ope}$(b)$ to acquire the robustness against rotation and translation of images.

The feature vectors extracted through the two-dimensional CNN operations are given to the three-dimensional CNN, which includes several convolutional layers and an upsampling layer before the output layer.
The internal procedure of the three-dimensional convolutional layer is illustrated in figure~\ref{fig:CNN_ope}$(c)$.
Similar to the two-dimensional operation, the input $X^{(\iota-1)}$ of size $L^{\prime}_1 \times L^{\prime}_2 \times L^{\prime}_3$ at the $(\iota-1)$th layer has $K^{\prime}$ channels.
The output can be then obtained through the convolutional operation with filter $h^{\prime{(\iota)}}$ of size $H^{\prime} \times H^{\prime} \times H^{\prime}$ with $L^{\prime}$ channels.
The output $X^{({\iota})}$ at the $\iota$th layer, pixel indices $(i,j,k)$ in a three-dimensional data, and filter index $m$, can be represented as
\begin{eqnarray}
    X^{(\iota)}_{ijkm} = {\varphi^{\prime}}\biggl({b_m^{\prime(\iota)}}+\sum^{L^{\prime}-1}_{l=0}\sum^{{H^{\prime}}-1}_{p=0}\sum^{{H^{\prime}}-1}_{q=0}\sum^{H^{\prime}-1}_{r=0}h_{pqrlm}^{\prime(\iota)} X_{i+p{{-G^{\prime}}},j+{q{-G^{\prime}}},k+r-G^{\prime},l}^{(\iota-1)}\biggr),
\end{eqnarray}
where {$G^{\prime}={\rm floor}(H^{\prime}/2)$,} 
$b^\prime_m$ is the bias, and $\varphi^\prime$ is an activation function, similar to the two-dimensional operation in equation~(\ref{eq:2DCNN}).
The upsampling layer illustrated in figure~\ref{fig:CNN_ope}$(d)$ plays a role of dimension extension.

The details of the proposed model are presented in table~\ref{tab:str_of_2D3DCNN}.
We use three velocity components ${\bm q}=\{u,v,w\}$ as the input and output attributes.
The features of input sectional data ${\bm q}_{\rm 2D}$ are extracted through two-dimensional convolutional layers and reduced in dimension by the max-pooling layer.
Since the pooling operations enable CNN models to acquire robustness against the variance of input data due to a decrease in spatial sensitivity~\cite{LBBH1998}, we here consider a maxpooling operation as noted as ``1st Max Pooling" in table~\ref{tab:str_of_2D3DCNN}.
Once the data shape is converted for the three-dimensional convolutional layers at the ``Reshape" layer, the data heads to the output layer.
The output ${\bm q}_{\rm 3D}$ is finally obtained via three-dimensional convolutional layers and upsampling.
Users can also consider the use of optimization tools for the parameter tuning inside CNNs, e.g., hyperopt~\cite{BYC2013} and Bayesian optimization~\cite{MMLMBL2019}.

\begin{table}[tbp]
    \centering
    \vspace{3mm}
    \caption{The network structure of the present 2D-3D CNN. The convolution layer is denoted as Conv.}
    \vspace{3mm}
    \label{tab:str_of_2D3DCNN}
    \begin{tabular}{ccc}\hline\hline
         Layer (filter size, \# of filters) & Data size & Activation\\ \hline
         Input & (256,128,$3\times n_{\rm section}$) & \\
         1st Conv2D (3,32) & (256,128,32) & ReLU\\
         2nd Conv2D (3,32) & (256,128,32) & ReLU\\
         1st Max Pooling    & (128,64,32) & \\
         3rd Conv2D (3,16) & (128,64,16) & ReLU\\
         4th Conv2D (3,20) & (128,64,20) & ReLU\\
         Reshape    & (128,64,20,1) & \\
         1st Conv3D (3,16) & (128,64,20,16) & ReLU\\
         2nd Conv3D (3,16) & (128,64,20,16) & ReLU\\
         1st Up-sampling    & (256,128,160,16) & \\
         3th Conv3D (3,32) & (256,128,160,32) & ReLU\\
         4th Conv3D (3,32) & (256,128,160,32) & ReLU\\
         5th Conv3D (3,3) & (256,128,160,3) & Linear\\ \hline\hline
    \end{tabular}
\end{table}

The 2D-3D CNN model $\mathcal{F}$ in the present study is trained in a supervised learning manner to obtain the three-dimensional data ${\bm q}_{\rm 3D}$ from the input sectional data ${\bm q}_{\rm 2D}$.
The training of the present model hence results in an optimization problem,
\begin{eqnarray}
    {\bm w} = {\rm argmin}_{{\bm w}}||{\bm q}_{\rm 3D}-{\cal F}({\bm q}_{\rm 2D}; {\bm w})||_2, \label{eq:CNNL2}
\end{eqnarray}
where ${\bm w}$ denotes the weights inside the present CNN.
For the construction of the present model, we take an early stopping criterion~\cite{prechelt1998} with 20 iterations to avoid an overfitting~\cite{BK2019book}.

\subsection{Super-resolution reconstruction with adaptive sampling}
\label{sec:method-as}

\begin{figure}[t]
    \vspace{0mm}
	\begin{center}
		\includegraphics[width=0.75\textwidth]{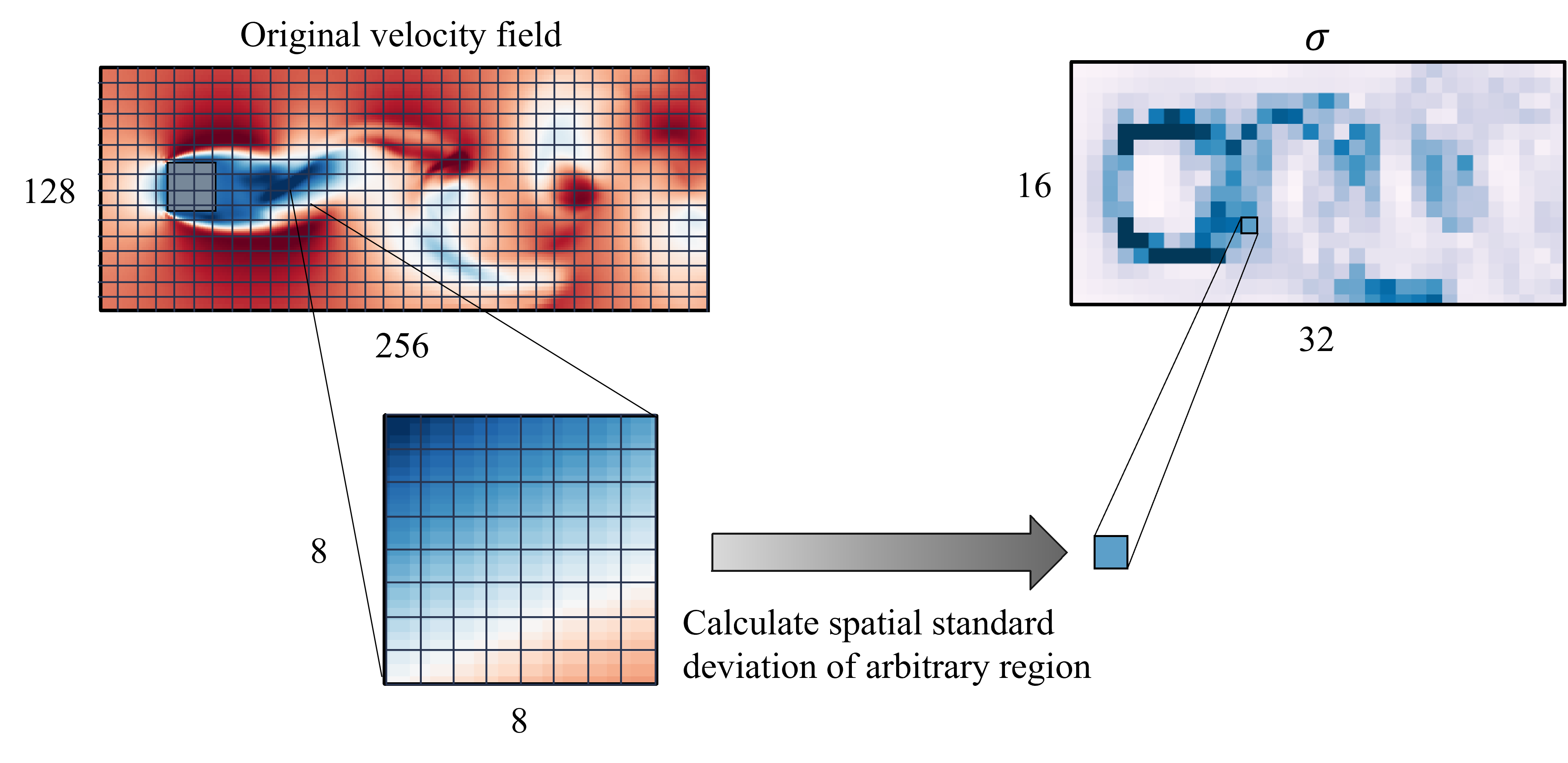}
		\caption{Generation of a standard deviation map from velocity data. An arbitrary area of $N_{sw}\times N_{sh}=8\times8$ is considered to calculate the spatial standard deviation.}
		\label{fig:sigma_map}
	\end{center}
\end{figure}

The current study also examines a combination of the present 2D-3D CNN and super-resolution analysis.
A machine learning model ${\cal G}$, which is distinct against the model for three-dimensional reconstruction ${\cal F}$, is utilized to reconstruct two-dimensional high-resolution sections ${\bm q}_{\rm 2D}^{\rm{HR}}$ from their two-dimensional low-resolution data ${\bm q}_{\rm 2D}^{\rm{LR}}$.
The super-resolved two-dimensional sections ${\bm q}_{\rm 2D}^{\rm{HR}}$ are then fed into the model for three-dimensional reconstruction ${\cal F}$.
The procedure can be expressed as 
\begin{eqnarray}
    {\bm q}_{\rm 2D}^{\rm HR} = {\cal G}({\bm q}_{\rm 2D}^{\rm LR};{\bm w}_{\cal G}), ~~{\bm q}_{\rm 3D} = {\cal F}({\bm q}_{\rm 2D}^{\rm HR};{\bm w}),
    \label{eq:sr}
\end{eqnarray}
where ${\bm w}_{\cal G}$ denotes the weights inside the super-resolution model ${\cal G}$.
This idea promotes more intense data compression than cases without utilizing super-resolution reconstruction.

To perform the super-resolution analysis, low-resolution data are prepared for the training phase.
They are usually generated through pooling operations which extract a representative value (max/average) from an arbitrary area~\cite{FFT2019a}.
With these pooling operations, the data can be downsampled uniformly over the field.
Although this is a simple and fast method to prepare low-resolution data, it is inefficient for some regions because the complexity largely varies in space.
To prepare low-resolution data efficiently, this study focuses on the spatial standard deviation in each arbitrary sub-domain to determine the local degree of downsampling by accounting for the `importance' of the data in a pooling process, referred to as {\it adaptive sampling}.

The detailed procedure of adaptive sampling is as follows.
We first divide a velocity snapshot into small subdomains of $N_{sw}\times N_{sh}$, where $N_{sw}$ and $N_{sh}$ correspond to the width and the height of the sub-domain, respectively, as shown in figure~\ref{fig:sigma_map}.
In this study, we use $\{ N_{sw}$, $N_{sh}\} = \{8,8\}$.
The spatial standard deviation of each sub-domain is then calculated.
Note that we do not take the standard deviation over time, but over each instantaneous snapshot to adapt the pooling procedure for each snapshot.
Thus, as shown in figure~\ref{fig:sigma_map}, the map of spatial standard deviation $\sigma$ can be generated with the size of $N_{\sigma,w}\times N_{\sigma,h}$, where $N_{\sigma,w}=N_x/N_{sw}$ and $N_{\sigma,h}=N_y/N_{sh}$, and $N_x \times N_y$ indicates the size of the entire cross-section.

\begin{algorithm}
\caption{Adapting pooling rate $\alpha$}        
\label{alg1}                          
\begin{algorithmic}[1]
\FOR {$i=1$ to $N_{\sigma,w}$}
\FOR {$j=1$ to $N_{\sigma,h}$}
\IF{$\sigma_{i,j}<\gamma_{th,1}$ and $\sigma_{i+1,j}<\gamma_{th,1}$ and $\cdots$ and $\sigma_{i+(\alpha/N_{sw}-1),j+(\alpha/N_{sh}-1)}<\gamma_{th,1}$}
\STATE $\alpha=\alpha_1$
\STATE ${\bm q}^{\rm LR}_{i,j}
={\bm q}^{\rm LR}_{i+1,j}
=\cdots
={\bm q}^{\rm LR}_{i+(\alpha/N_{sw}-1),j+(\alpha/N_{sh}-1)}
=\frac{1}{\alpha^2}\sum_{p,s}^{\alpha}({\bm q}^{\rm HR}_{psij})$
\vspace{5mm}
\ELSIF{$\gamma_{th,1}<\sigma_{i,j}<\gamma_{th,2}$ and $\gamma_{th,1}<\sigma_{i+1,j}<\gamma_{th,2}$ and $\cdots$ \\and
$\gamma_{th,1}<\sigma_{i+(\alpha/N_{sw}-1),j+(\alpha/N_{sh}-1)}<\gamma_{th,2}$
}
\STATE $\alpha=\alpha_2$
\STATE ${\bm q}^{\rm LR}_{i,j}
={\bm q}^{\rm LR}_{i+1,j}
=\cdots
={\bm q}^{\rm LR}_{i+(\alpha/N_{sw}-1),j+(\alpha/N_{sh}-1)}
=\frac{1}{\alpha^2}\sum_{p,s}^{\alpha}({\bm q}^{\rm HR}_{psij})$
\vspace{5mm}
\ELSIF{$\gamma_{th,2}<\sigma_{i,j}$}
\STATE $\alpha=\alpha_3$
\STATE ${\bm q}^{\rm LR}_{i,j}
=\frac{1}{\alpha^2}\sum_{p,s}^{\alpha}({\bm q}^{\rm HR}_{psij})$
\vspace{5mm}
\ENDIF
\ENDFOR
\ENDFOR
\end{algorithmic}
\end{algorithm}
\vspace{5mm}

Using the standard deviation map $\sigma$, we prepare the adaptive sampling-based low-resolution data with the local pooling rate $\alpha$.
The adaptive sampling rate $\alpha$ decides how strong 
features we retain following the standard deviation map which represents the strength of features (i.e., color in figure~\ref{fig:sigma_map}) on each snapshot.
The detailed procedure of adaptive sampling is summarized in Algorithm~\ref{alg1}.
Through the algorithm, we obtain an adaptive-sampled low-resolution data ${\bm q}^{\rm LR}$ by setting a corresponding high-resolution data ${\bm q}^{\rm HR}$, its standard deviation map $\sigma$ and a threshold value $\gamma_{th}$.

For demonstration, we will compare the performance of the super-resolution reconstruction with the average- and the present adaptive sampling in section~\ref{sec:adaptive}.
Since the adaptive sampling-based low-resolution data can hold more condensed spatial information than the average one, the expected result is that the adaptive sampling-based method reconstructs better than the average sampling-based method when the number of data points on the low-resolution field is the same.
To confirm this point, we prepare the adaptive-sampled data using the algorithm for each velocity component so that 
the number of data points for each velocity component becomes nearly equal to that of the average-pooled data (i.e., the number of grid points).
Because the structural distribution and its complexity are different among the three velocity components, we have to choose the threshold value $\gamma_{th}$ individually for each variable.
We set the threshold value $\gamma_{th}$ to $\{ \gamma_{th,1u},\gamma_{th,2u},\gamma_{th,1v},\gamma_{th,2v},\gamma_{th,1w},\gamma_{th,2w} \}=\{0.08, 0.1,0.06, 0.08, 0.05,0.07 \}$ for each velocity component.
The pooling rates are commonly set for all velocity components as $\{\alpha_1,\alpha_2,\alpha_3\}=\{32,16,8\}$.

\begin{figure}[t]
	\vspace{0mm}
	\begin{center}
		\includegraphics[width=0.7\textwidth]{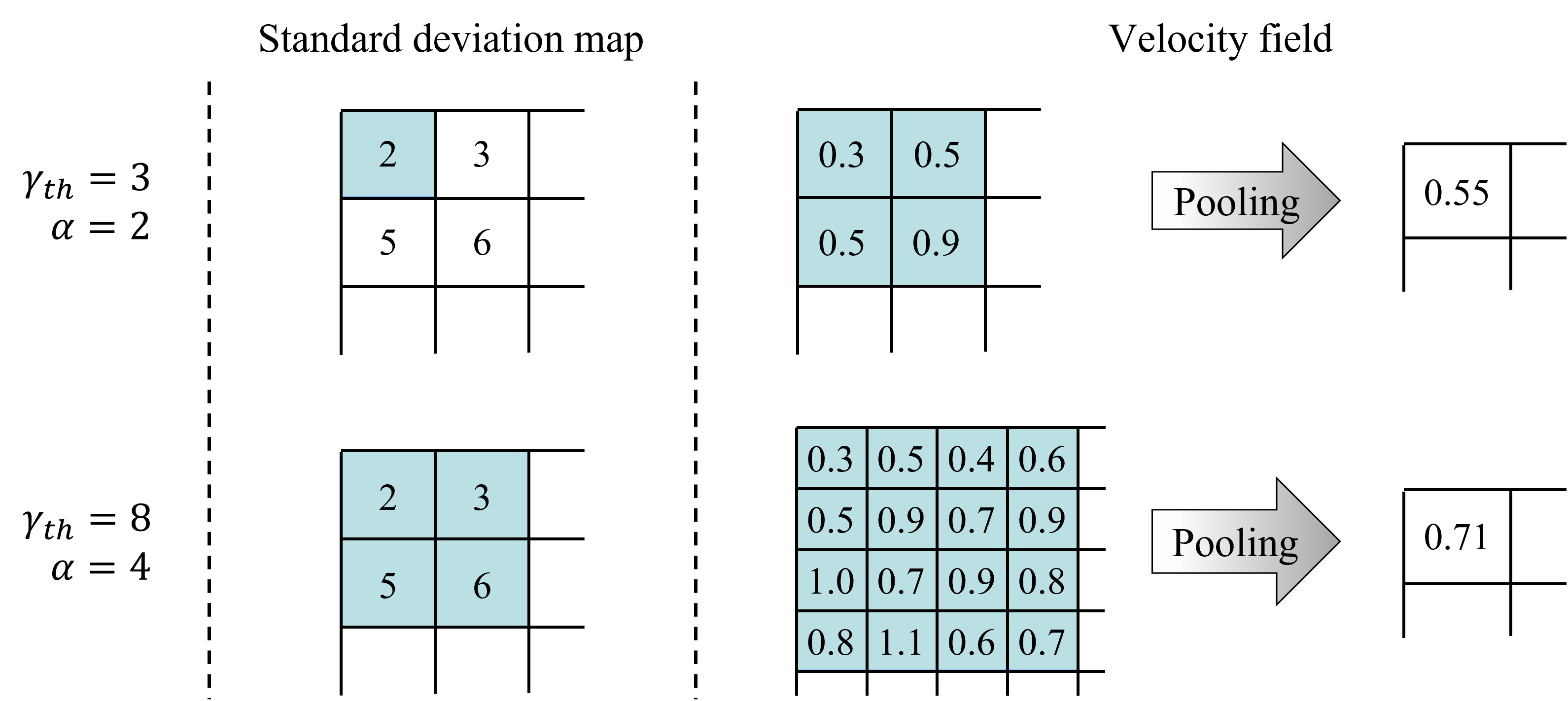}
		\caption{Adaptive sampling operation in this study.}
		\label{fig:adpt_pool}
	\end{center}
\end{figure}

An illustrative example of the adaptive sampling operation is shown in figure~\ref{fig:adpt_pool}.
For example, let us focus on the case with $\gamma_{th}=3$ and the pooling rate $\alpha=2$.
In this case, the pink region in the standard deviation map (where $\sigma=2<\gamma_{th}$) is only treated as the target region for the pooling operation.
The region in the velocity field corresponding to the target region (having $\{0.3, 0.5, 0.5, 0.7\}$ with the size of $2\times2$) is then downsampled by using the average pooling.
Consequently, we obtain 0.25 as the representative value as shown in figure~\ref{fig:adpt_pool}.
In sum, the downsampled velocity field at the region can be expressed with the data size of $1/\alpha^2\ (=25\%)$ of the original data size.
Similarly, for the case with $\gamma_{th}=10$ and $\alpha=4$ illustrated in the lower part in figure~\ref{fig:adpt_pool}, the target pooling region is expanded to the size of $4\times4$ having $\{2,4,6,8\}$.
In this case, the velocity field can be downsampled by $1/\alpha^2\ (=6.25\%)$ of the original size.
In other words, the portions having small standard deviations are discarded through the present operation --- in turn, we can retain the region with a larger standard deviation representing the strong feature of the velocity field.

\begin{figure}[t]
    \vspace{0mm}
	\begin{center}
		\includegraphics[width=0.99\textwidth]{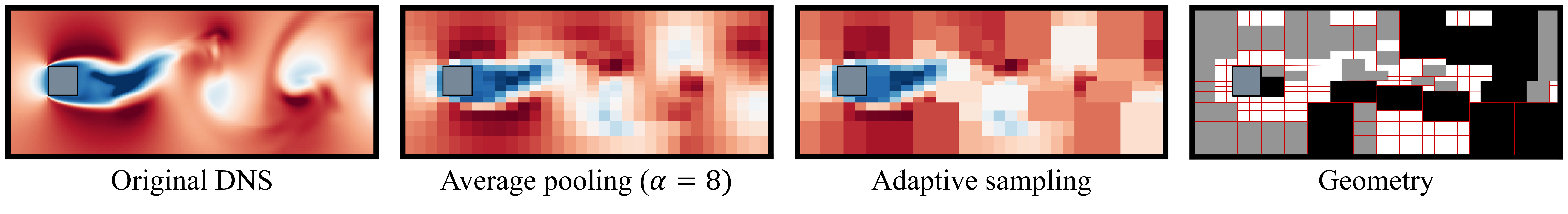}
		\caption{Application of the adaptive sampling to the present velocity field.
		For comparison, the standard average pooling with $\alpha=8$ is also shown.
		The colors in the geometry image indicate the pooling rate $\alpha$; white represents $\alpha=8$, gray represents $16$ and black represents $32$.}
		\label{fig:compare_velocity_field}
	\end{center}
\end{figure}

The application of the aforementioned idea to the velocity field is presented in figure~\ref{fig:compare_velocity_field}.
As expected, the region with the higher spatial standard deviation corresponding to the shear layer of wake provides a higher resolution than that with the less spatial variation region.
Note that we have the variation for the domain size among the same pooling rate since we consider the non-uniform grid in the $y$ direction.
In section~\ref{sec:adapsuper_results}, we will demonstrate the proposed adaptive sampling in handling fluid flow data.

\section{Results and Discussion}
\label{sec:result}

\begin{figure}[t!]
	\vspace{0mm}
	\begin{center}
		\includegraphics[width=0.84\textwidth]{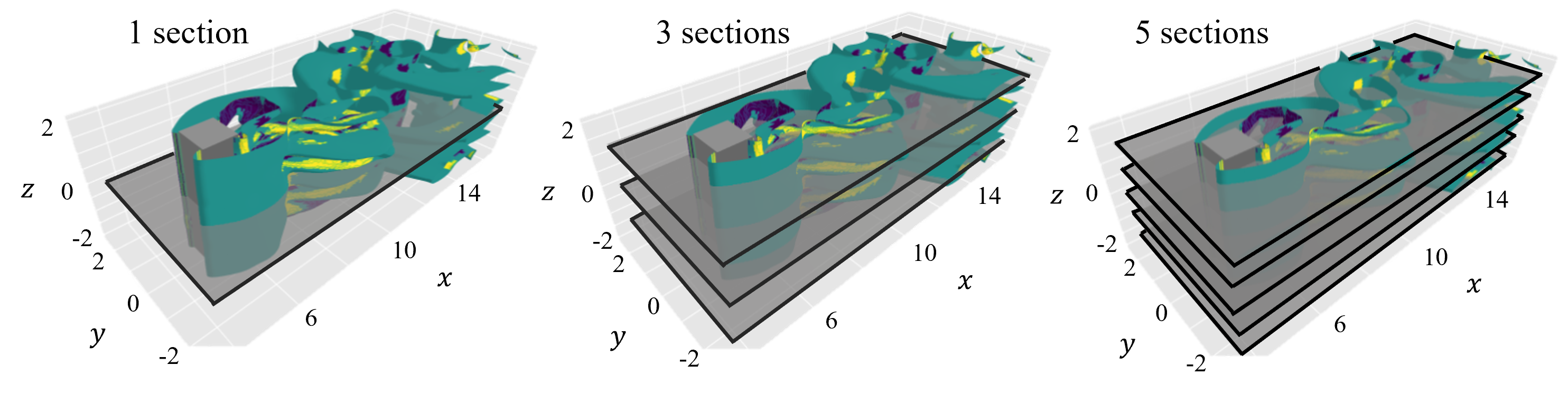}
		\caption{Schematics of input sections.}
		\label{fig:input}
	\end{center}
	\vspace{0mm}
	\begin{center}
		\includegraphics[width=0.9\textwidth]{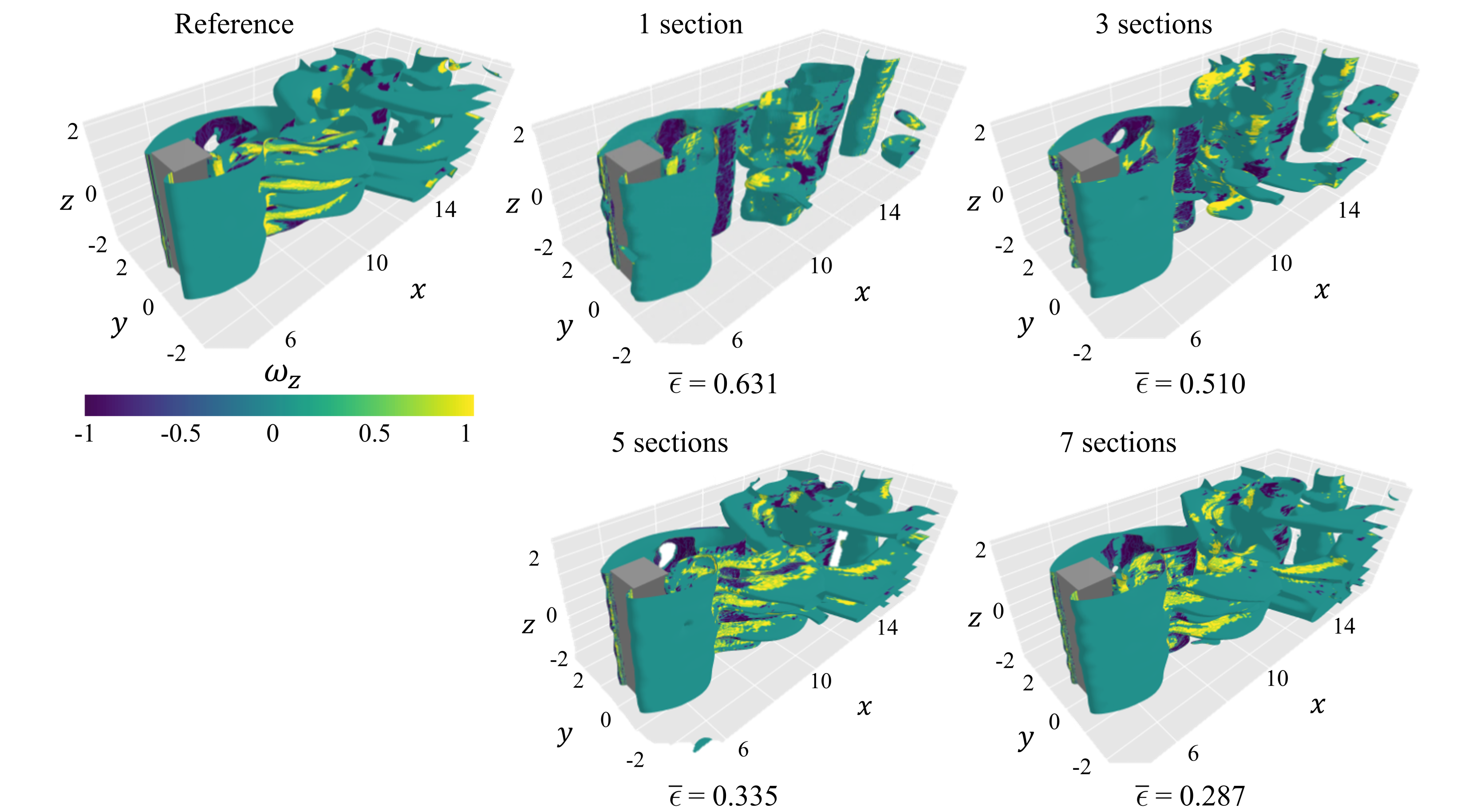}
		\caption{Convolutional neural network-based three-dimensional velocity field reconstruction fields using 1,\ 3,\ 5 and 7 cross-sections (vortical structures identified by $\lambda _2 = -0.001$ and colored by the spanwise vorticity $\omega_z$). The values underneath each subfigure indicate the ensemble $L_2$ error norm over all velocity attributes and structural similarity.}
		\label{fig:reconstructed_field}
	\end{center}
\end{figure}

We first assess the three-dimensional fields reconstructed from the high-resolution two-dimensional velocity field data prepared by DNS (section~\ref{sec:three-twohigh}). 
We also discuss the possibility of more efficient data compression and reconstruction using a combination with the super-resolution concept (section~\ref{sec:adapsuper_results}).

\subsection{Three-dimensional reconstruction from high-resolution two-dimensional sections}
\label{sec:three-twohigh}

In this section, we reconstruct a three-dimensional velocity field from two-dimensional high-resolution cross sections.
Four different numbers of input sections $n_{\rm{section}}=\{1,3,5,7\}$ are considered as illustrated in figure~\ref{fig:input}.
We place the baseline input section corresponding to $n_{\rm section}=1$ at the middle of the $z$ direction.
Three or more input sections are added with equal intervals in the $z$ direction.
The locations of the input section are fixed over time and all velocity attributes in all cases.

The reconstructed three-dimensional fields are visualized in figure~\ref{fig:reconstructed_field}. 
We use a $\lambda_2$ vortex criterion~\cite{JH1995} with $\lambda_2=-0.001$ to identify the vortical structures.
The use of more input sections provides better wake reconstruction.
The present method can estimate the large structure in the flow field even from one cross-section in a qualitative manner.
The fields reconstructed from five or more input sections have detailed three-dimensional structures and are in agreement with the reference data.
This can also be observed through the ensemble $L_2$ error norms over the velocity attributes $\overline{\epsilon} = {||{\bm{q}}_{\rm{{DNS}}}-{\bm{q}}_{\rm{ML}}||_2}/{||{\bm {q}} \prime_{\rm{DNS}}||_2}$, where $(\cdot)^\prime$ denotes fluctuations of the velocity attributes, shown underneath each subfigure.

\begin{figure}[t]
	\vspace{0mm}
	\begin{center}
		\includegraphics[width=0.9\textwidth]{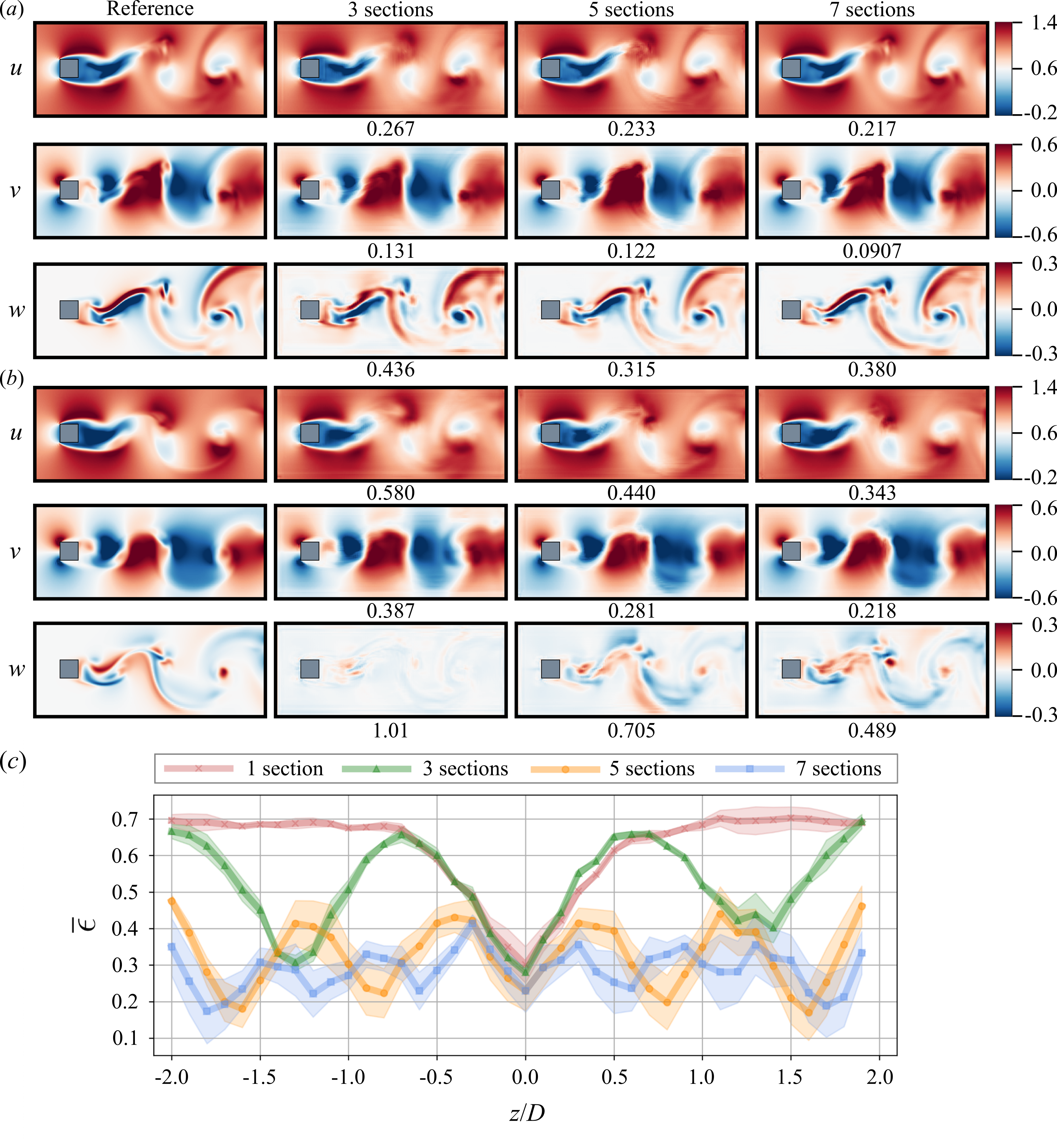}
		\caption{Details on convolutional neural network-based three-dimensional data reconstruction from two-dimensional sections. Reconstructed cross-sectional velocities at $(a)$ $z/D=0.0$ and $(b)$ $z/D=-2.0$. 
		The values underneath the velocity contours indicate the $L_2$ error norm $\overline{\epsilon}$.
		$(c)$ Dependence of $L_2$ error norm on the location in the $z$ direction. The confidence intervals represent the standard deviation among three-fold cross-validation.}
		\label{fig:2D-3D-contour}
	\end{center}
\end{figure}

To further assess the current reconstruction, the estimated cross-sectional velocity fields at 
$z/D = 0.0$, which is the same position as the input section, are shown in figure~\ref{fig:2D-3D-contour}$(a)$.
The present models reconstruct the flow field with reasonable accuracy in terms of both the visualization and the $L_2$ error norms.
Especially, the reconstructed fields with $n_{\rm section}=7$ capture the fine structures, which is almost indistinguishable from the reference.
The velocity fields estimated at $z/D = -2.0$, which is relatively far from the input sections for all cases, are also depicted in figure~\ref{fig:2D-3D-contour}$(b)$.
Although the model trained $n_{\rm section}=3$ can hardly estimate the $w$ component, the other reconstructed fields are in qualitative agreement with the reference even at the edge of the reconstruction domain.

The discussion above focused on the results in the particular sections.
Let us then evaluate the dependence of the $L_2$ error norm and standard deviation among the three-fold cross-validation on the location in the $z$ direction in figure~\ref{fig:2D-3D-contour}$(c)$.
There are some valleys of the $L_2$ error near the input section because the ML is likely good at estimating fields near the input section.
The error trends with a large number of input sections are entirely lower than those with smaller numbers.

\begin{figure}[t]
	\vspace{0mm}
	\begin{center}
		\includegraphics[width=0.9\textwidth]{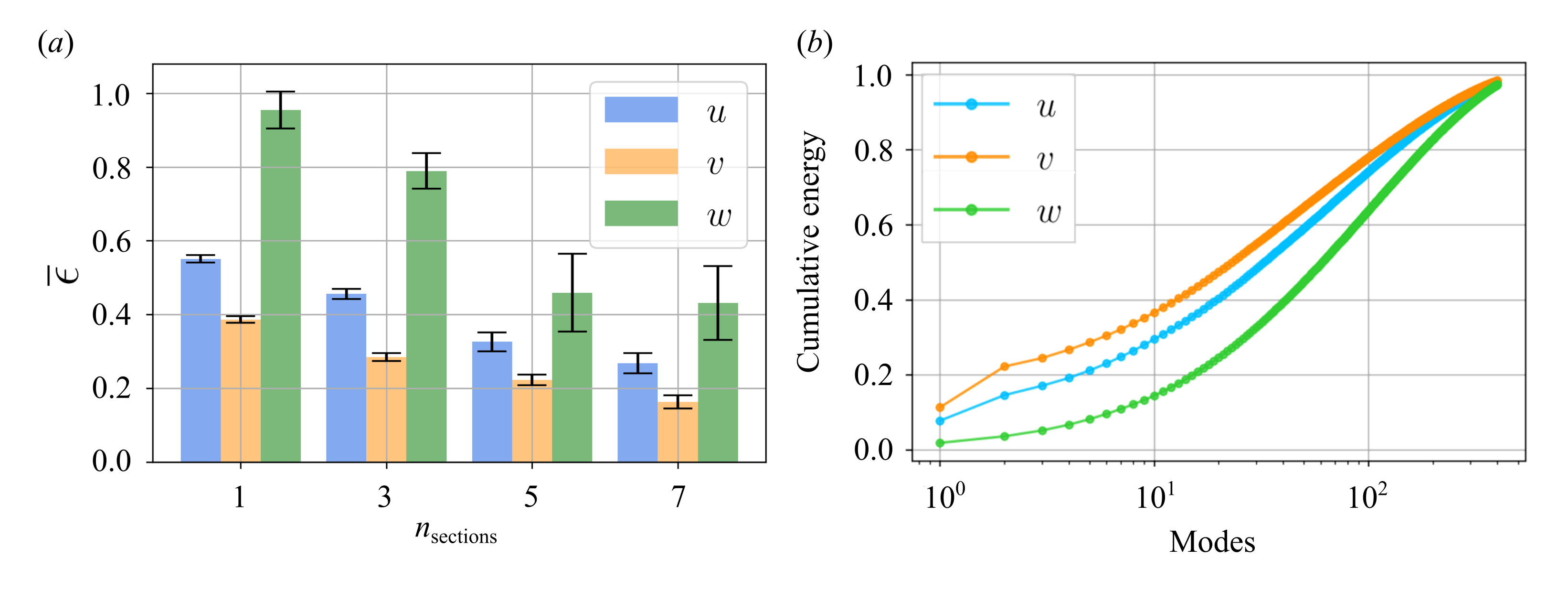}
		\caption{$(a)$ Dependence of $L_2$ error norm on the velocity attributes. Error bars represent the variance over the present cross validation. $(b)$ Singular value spectra of velocity fields.}
		\label{fig:energy_spectrum}
	\end{center}
\end{figure}

We also summarize the relationship between the ensemble $L_2$ error norm, among the three-fold cross validations, over the domain for each velocity attribute and the number of input sections in figure~\ref{fig:energy_spectrum}$(a)$.
The error bars represent the standard deviation among the three-fold cross validation.
Similar to figures~\ref{fig:reconstructed_field}, and~\ref{fig:2D-3D-contour}, the larger $n_{\rm section}$, the better reconstruction.
We also observe a significant difference in the estimation accuracy among velocity attributes --- the $L_2$ error of the $v$ component reports much lower values than that of the $w$ component.

The difficulty for the $w$ component coincides with P\'erez et al.~\cite{PCV2020} who used a spatio-temporal Koopman decomposition for three-dimensional wake reconstruction around a square cylinder at $Re_D=280$ from two-dimensional data.
It can also be verified from the aspect of the data complexity of each attribute.
We take the singular value decomposition (SVD) for snapshots of each velocity attribute, as presented in figure~\ref{fig:energy_spectrum}$(b)$.
The rate of energy accumulation for the $w$ component is much slower than that of the other components.
This example analysis implies that we may be able to relate the assessment of machine-learning-based estimation with linear analyses such as SVD~\cite{MFZF2020}.

\begin{figure}[t]
	\vspace{0mm}
	\begin{center}
		\includegraphics[width=0.95\textwidth]{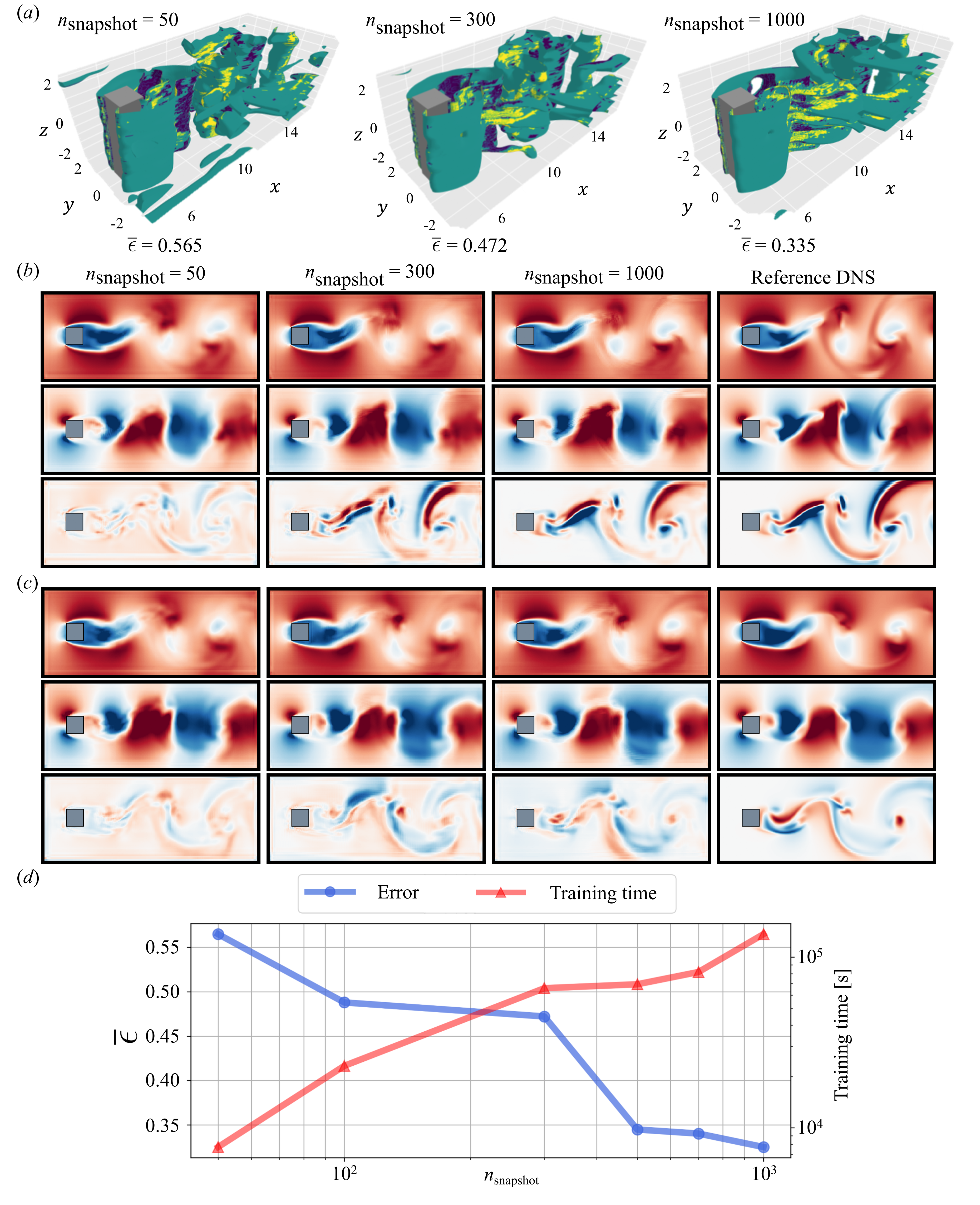}
		\caption{
		Dependence of estimation ability on the number of training snapshots and training time. $(a)$ Three-dimensional reconstructed fields. Reconstructed cross-sectional velocities at $(b)$ $z/D = 0.0$ and $(c)$ $z/D = -2.0$ are shown. $(d)$ Relationship between reconstruction error and training time.}
		\label{fig:numsnap1}
	\end{center}
\end{figure}

We examine the dependence of the estimation accuracy on the number of training snapshots, as presented in figure~\ref{fig:numsnap1}.
We use $n_{\rm section}=5$ for this investigation.
The time interval among snapshots is fixed for all covered $n_{\rm snapshot}$.
Hence, the length of reference time over training data increases with $n_{\rm snapshot}$ while the temporal density among snapshots is kept the same for all models.
As mentioned earlier, we use 100 snapshots as the test data excluded from the training/validation data.

The error decreases with increasing the number of training snapshots.
Interestingly, the present model trained with $n_{\rm snapshot}=50$ is able to capture large-scale structures of velocity $u$ and $v$ components even at $ z/D = -2.0$, which is far from the input sections,
although the field of $w$ component is blurry.
This difference among the velocity attributes can also be explained by their complexities observed in figure~\ref{fig:energy_spectrum}$(b)$.
With $n_{\rm snapshot}\ge300$, the present model can achieve a reasonable reconstruction.
We also investigate the computational cost for constructing the present models, shown as the red curve in figure~\ref{fig:numsnap1}$(d)$.
It takes approximately 38 hours with the NVIDIA TESLA V100 graphics processing unit (GPU) for the case with $n_{\rm snapshot}=1000$.
Note that the increase in computational costs does not show a linear behavior because of the use of early stopping as mentioned above.
Care should be taken about the trade-off relationship between the computational cost and the estimation accuracy depending on their requirements.

The aforementioned results suggest the applicability of the present model to particle image velocimetry, e.g., estimation of the whole flow field from two-dimensional sections that can only be measured due to experimental constraints.
Here, we also assess the robustness of the present model against noisy input~\cite{NAKAMURA2022108997}.
To determine the magnitude of given noise, we use the signal-noise ratio (SNR) 
as ${\mathrm {SNR}} = \sigma^2_{\rm sensor}/\sigma^2_{\rm noise}$, where $\sigma^2_{\rm sensor}$ and $\sigma^2_{\rm noise}$ represent the variance of data and the variance of noise, respectively.
We test six cases of noise magnitudes as $1/{\mathrm{SNR}}=\{0.001,0.005,0.01,0.05,0.1,0.5\}$.
We use Gaussian noise for the noisy-input generation.
The mean value of Gaussian noise is $\mu=0$ and the standard deviation is $\sigma_{\rm noise} = \sigma_{\rm sensor} / {\sqrt{\rm SNR}}$.

\begin{figure}[t]
	\vspace{0mm}
	\begin{center}
		\includegraphics[width=0.95\textwidth]{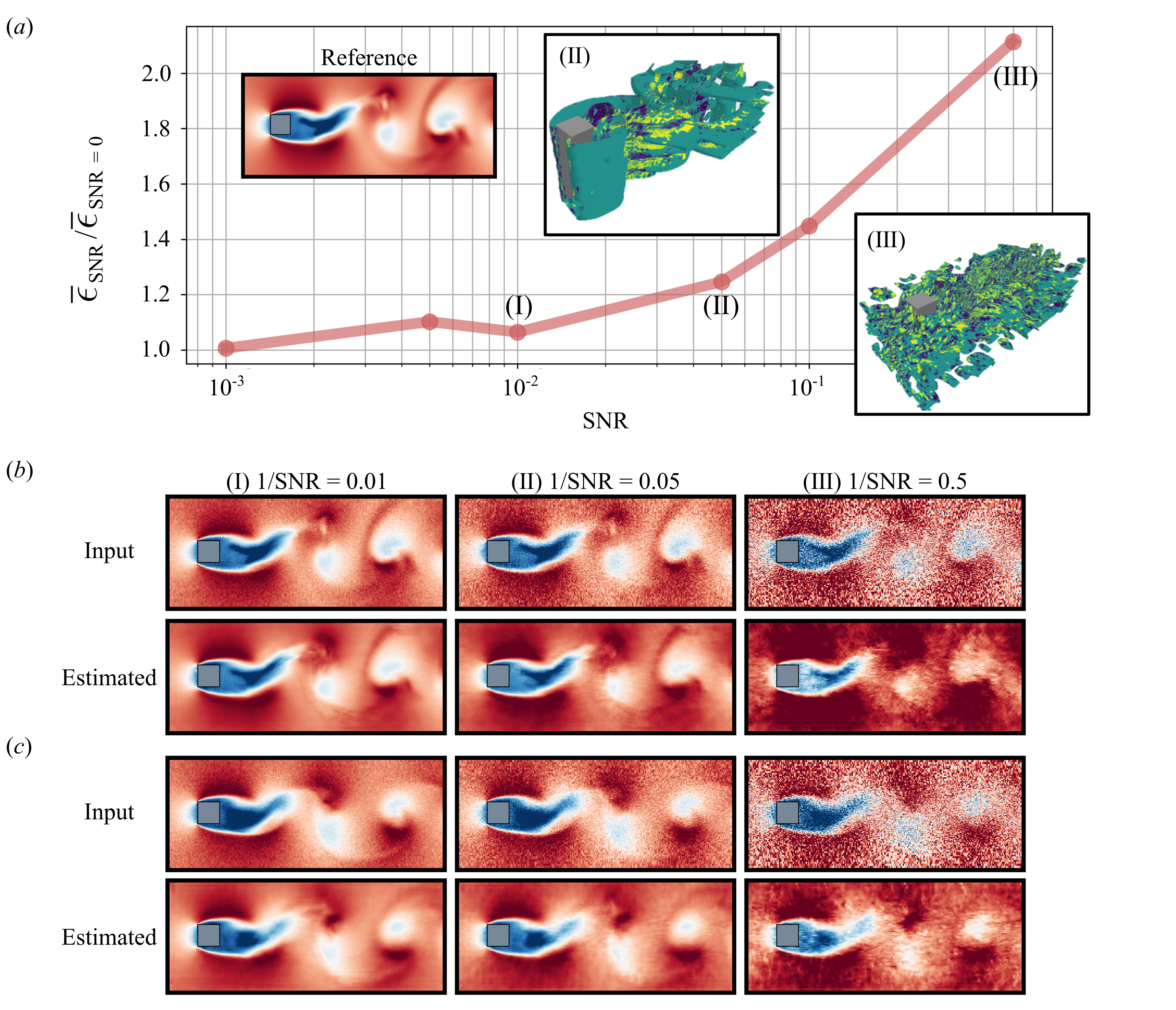}
		\caption{Robustness of the present model against noisy input. $(a)$ Relationship between reconstruction error and magnitude of noise. The error is normalized by that without noise $\overline{\epsilon}_{\rm SNR=0}$. Reconstructed cross-sectional velocity of $u$ component at $(b)$ $\Delta z/D = 0.0$ and $(c)$ $\Delta z/D =-2.0$.}
		\label{fig:noise}
	\end{center}
\end{figure}

The investigation for the robustness against noisy input data is summarized in figure~\ref{fig:noise}.
We use $n_{\rm section}=5$.
Up to $1/{\mathrm{SNR}}=0.01$, the estimation error is almost the same as that without noise.
The estimated fields also show no significant difference against the reference DNS.
However, with $1/{\mathrm{SNR}}=0.05$ or more, the estimated fields do not include fine structures in addition to the overestimation of their velocity magnitudes.
The aforementioned trend is consistent between the sections at the input position and $z/D=-2.0$, which is far from the input section.

\begin{table}[tbp]
    \centering
    \vspace{3mm}
    \caption{Comparison of required data storage for saving flow data and ML model.}
    \vspace{3mm}
    \label{tab:comparing-storage}
    \begin{tabular}{ccccc}
    \hline\hline
          & Sections to save & $\overline{\epsilon}$ & Data (GB) & ML model (MB)\\ \hline
         DNS & 160 & -- & 566.6 & --\\ \hline
         \multirow{4}{*}{2D-3D CNN} & 7 & 0.287 & 24.8 & 1.89\\
         & 5 & 0.335 & 17.7 & 0.916\\
         & 3 & 0.510 & 10.6 & 0.896\\
         & 1  & 0.631 & 3.54 & 0.876\\ \hline 
         \hline
    \end{tabular}
\end{table}

The actual sizes of data storage required for the DNS and the proposed method are compared in table~\ref{tab:comparing-storage}.
While approximately 570~GB is required to save 1000 snapshots of the DNS data, our method requires only 3.5--25~GB in total, which is a significant storage saving.
If one is able to transfer data to the collaborator at 1 Giga bit per second, the transfer time can be reduced from more than 1 hour to less than 4 minutes in this specific example.
Although the reconstruction ability and data storage are in a trade-off relationship, users can choose the compression ratio for saving the data depending on their requirements.
Moreover, the advantage of the method can also be seen in the computational time.
The trained model estimates a single snapshot taking $2.69 \times 10^{-1}$ seconds using GPU, which is significantly shorter than the integration time for DNS.
These comparisons imply that once the model is trained, the flow data can be obtained with significantly shorter computational time using only several sections.

\begin{figure}[t]
	\vspace{0mm}
	\begin{center}
		\includegraphics[width=1\textwidth]{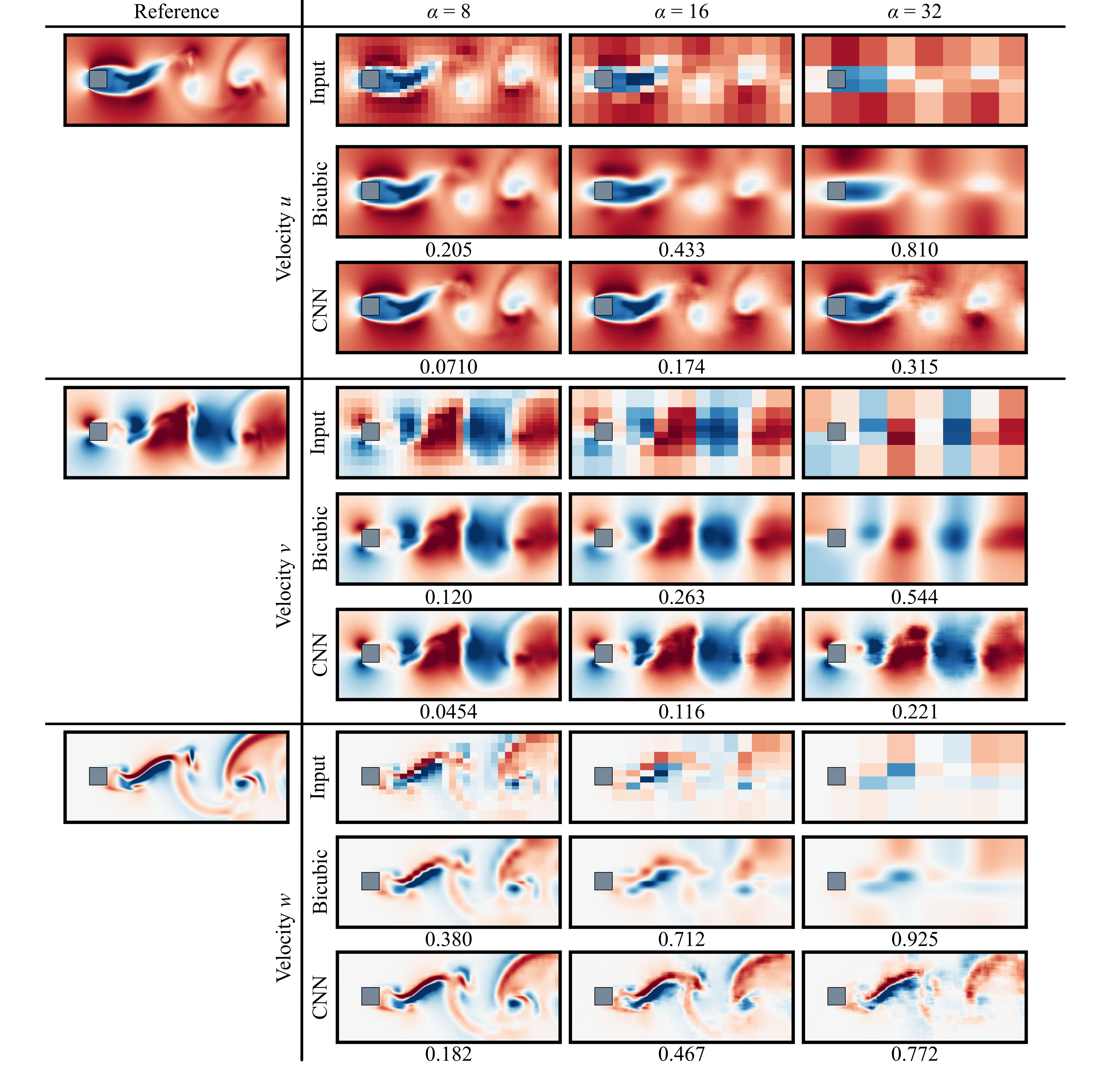}
		\caption{Super-resolution reconstruction for two-dimensional velocity fields. The values underneath the contours indicate the $L_2$ error norm $\overline{\epsilon}$. The contour level for each velocity attribute is the same as that used in figure \ref{fig:2D-3D-contour}.}
		\label{fig:2DSR}
	\end{center}
\end{figure}

\subsection{Super-resolution aided efficient data handling}
\label{sec:adapsuper_results}

As introduced in section~\ref{sec:method-as}, we seek an extensional possibility of the present method.
To estimate three-dimensional data from a low-dimensional two-dimensional sectional dataset, we consider a combination with super-resolution analysis for more efficient data handling.
Super resolution reconstructs high-resolution data from its low-resolution data.
Fukami et al.~\cite{FFT2019a,FFT2020b}~recently proposed a CNN-based super-resolution reconstruction for turbulent flows.
The procedure of the combination of the present three-dimensional reconstruction with the super-resolution analysis is expressed in what follows:
\begin{enumerate}
    \item Two-dimensional high-resolution flow fields ${\bm q}_{\mathrm {2D}}^{\rm HR}$ are reconstructed from low-resolution counterparts ${\bm q}_{\mathrm {2D}}^{\mathrm {LR}}$ through the super-resolution such that ${\bm q}_{\mathrm {2D}}^{\mathrm {HR}}={\cal G}({\bm q}_{\mathrm {2D}}^{\mathrm {LR}})$;
    \item The super-resolved two-dimensional sections ${\bm q}_{\mathrm {2D}}^{\mathrm {HR}}$ are then utilized as the input for the model of three-dimensional reconstruction such that ${\bm q}_{\rm 3D}={\cal F}({\bm q}_{\mathrm {2D}}^{\mathrm {HR}})$.
\end{enumerate}
These procedures allow us to save only low-resolution sectional flow fields ${\bm q}_{\mathrm {2D}}^{\rm LR}$ to represent the three-dimensional field ${\bm q}_{\rm 3D}$, as expressed in equation~\ref{eq:sr}.
For the machine learning model ${\cal G}$ for super-resolution analysis, we use a two-dimensional CNN which contains 7 layers and $5 \times 5$ filters with ReLU activation function.
Since the input low-resolution data is resized to the high-resolution image size before feeding into the model, the pooling or upsampling layers are not considered for the construction of model $\cal G$.

The separate construction of the two-dimensional super-resolution model $\cal G$ and the three-dimensional reconstruction model $\cal F$ provides flexibility for users.
Depending on the complexity of data that users handle, the coarseness of low-resolution input for the super-resolution model $\cal G$ can be modified while keeping the model $\cal F$.
This also allows us to skip training the model $\cal F$ again, which is prohibitively expensive due to the three-dimensionality.

In what follows, we apply a normal super-resolution analysis with a conventional pooling operation (section~\ref{sec:typicalSR}).
Moreover, we also examine the possibility of additional data saving with the super-resolution analysis by considering {\it adaptive sampling} (section~\ref{sec:adaptive}).

\begin{table}[t]
    \centering
    \caption{Computational costs for super-resolution analysis. The computational times for training of the present two-dimensional CNN-based super-resolution model with 1000 snapshots and reconstruction of one snapshot by the bicubic and the CNN are reported. We here use the low-resolution data with $\alpha=8$.}
    \label{tab:comp}
    \vspace{3mm}
    \begin{tabular}{lcc}
    \hline \hline
        &Bicubic &CNN\\ \hline 
         Training (h)& - & $22.2$\\
         Reconstrcution (s) &$5.53 \times 10^{-1}$& $5.41 \times 10^{-1}$\\
         \hline \hline
    \end{tabular}
\end{table}
\begin{figure}
	\vspace{0mm}
	\begin{center}
		\includegraphics[width=0.92\textwidth]{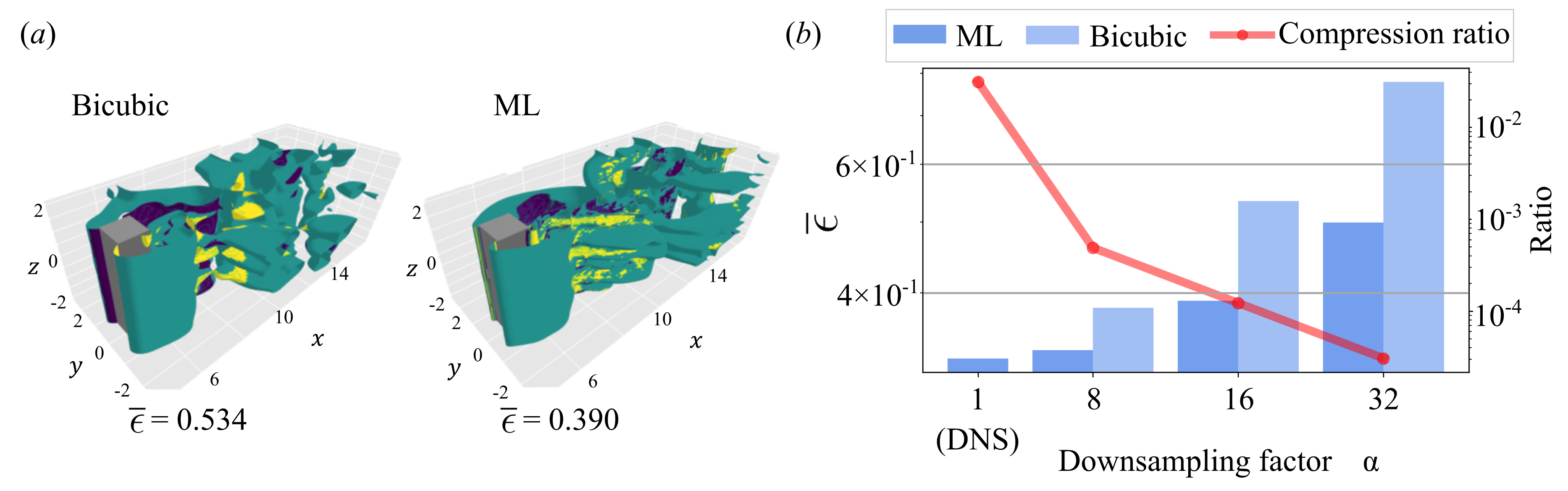}
		\caption{Three-dimensional reconstruction from the super-resolved two-dimensional data. $(a)$ The reconstructed fields from the super-resolved sectional data by bicubic interpolation (left) and machine learning (right). Values underneath the three-dimensional isosurfaces indicate the $L_2$ error norm and structural similarity. $(b)$ The relationship among the $L_2$ error norm, the compression ratio, and the downsampling factor $\alpha$.}
		\label{fig:3DSRave}
	\end{center}
\end{figure}

\subsubsection{Use of typical super-resolution}
\label{sec:typicalSR}

Before combining with the present 2D-3D CNN, let us examine super-resolution reconstruction of two-dimensional velocity fields itself.
To generate low-resolution data in this section, we use an average pooling which extracts a mean value in an arbitrary area, such that
\begin{flalign}
{\bm q}_{ij}^{\mathrm {LR}}=\frac{1}{\alpha^2}{\sum_{p,s}^{\alpha}({\bm q}_{psij}^{\rm HR})},
\end{flalign}
where $\alpha$ denotes a downsampling factor.
We consider three cases of downsampling factor $\alpha=8,~16,$ and $32$.
We also compare our reconstruction to bicubic interpolation~\cite{keys1981cubic}, a conventional method for super-resolution analysis.

We show the present two-dimensional super-resolution analysis of the velocity fields in figure~\ref{fig:2DSR}.
As an overall trend, the super-resolution reconstruction is superior to the bicubic interpolation in terms of estimation accuracy with all downsampling factors, as reported in~\cite{FFT2019a}.
Although the bicubic interpolation exhibits reasonable reconstruction with $\alpha=8$, this trend can clearly be seen with $\alpha=16$ and 32.
The difference among velocity attributes is caused by their complexities as presented in figure~\ref{fig:energy_spectrum}.

We then compare the computational costs for training the machine-learning model of super-resolution analysis and reconstructing a high-resolution field in table~\ref{tab:comp}.
The CNN can reconstruct a flow field as fast as the bicubic interpolation although it takes approximately 22 hours to train the machine learning model on the NVIDIA TESLA V100 graphics processing unit (GPU) with $n_{\rm snapshot}=1000$.
However, we emphasize that the present machine-learned model can be used while achieving a better reconstruction with a reasonable estimation cost, once the training is successfully performed.

The super-resolved two-dimensional sectional data are then fed into the 2D-3D CNN.
We consider both the machine-learning-based super-resolution and bicubic interpolation for the input.
For the three-dimensional reconstruction model ${\cal F}$, we use $n_{\mathrm{section}}=5$.
Three-dimensional reconstructed fields from the super-resolved input with $\alpha=16$ are shown in figure~\ref{fig:3DSRave}$(a)$.
The field reconstructed from the input with bicubic interpolation is very blurry and does not contain fine structures, especially in the wake region.
In contrast, the reconstructed field from the machine-learning-based super-resolved data is in reasonable agreement with reference DNS.
In other words, the accuracy of input super-resolved fields affects directly the reconstruction accuracy of the three-dimensional flow field.

We assess the relationship between the reconstruction $L_2$ error and data compression ratio in each case of $\alpha$, in figure~\ref{fig:3DSRave}$(b)$.
The advantage of the use of machine-learning-based super-resolution analysis can also be observed with the other downsampling factor.
There is almost no difference between the cases with sectional data with $\alpha=8$ and original DNS data, which can achieve approximately 1/2000 compression against the original data as presented as the red curve in figure~\ref{fig:3DSRave}$(b)$.
Users can choose an appropriate factor by caring about the trade-off relationship among the reconstruction error, the computational storage, and the training cost for machine-learning models.

\subsubsection{Combination with adaptive sampling}
\label{sec:adaptive}

\begin{figure}[t]
	\vspace{0mm}
	\begin{center}
		\includegraphics[width=0.87\textwidth]{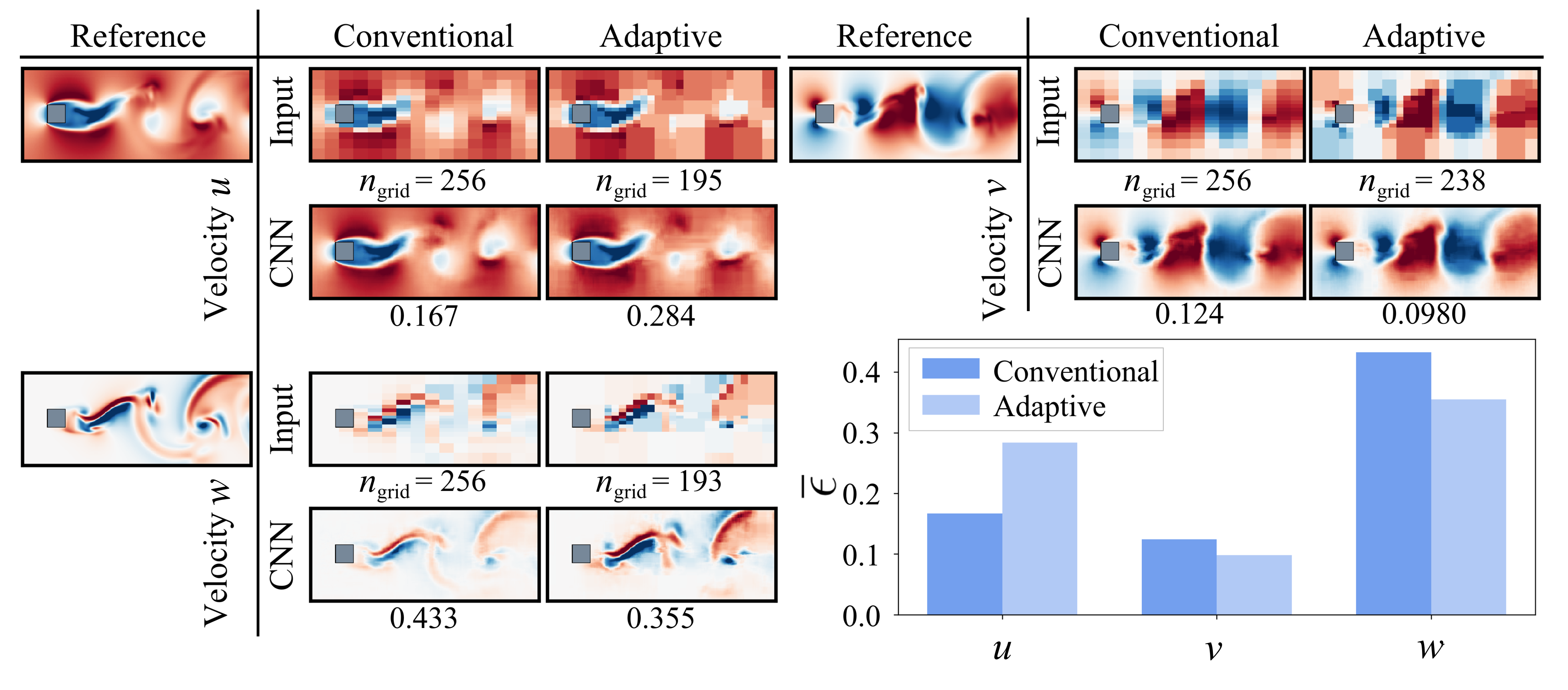}
		\caption{Super-resolution reconstruction from adaptive sampled coarse data. The values underneath the contours of input indicate the number of grid points. The values below the contours indicate the $L_2$ error norm. The relationship between the sampling method and the $L_2$ error norm depending on the velocity attributes is also shown.}
		\label{fig:2DSRadap}
	\end{center}
\end{figure}

As introduced in section~\ref{sec:method-as}, we also aim to achieve more efficient data handling with adaptive sampling.
The adaptive sampling generates low-resolution data based on the difference in the `importance' of each arbitrary region.
We present the adaptive sampling-based two-dimensional super-resolution analysis in figure~\ref{fig:2DSRadap}.
The considered downsampling factor $\alpha$ of conventional pooling in figure~\ref{fig:2DSRadap} is changed from the discussion above --- $\alpha_x=16$ in the streamwise direction and $\alpha_y=8$ in the normal direction --- to align the number of grid points with adaptive sampling $n_{\rm grid,~adaptive}$ as possible as we can.
The down-sampled images of average pooling and adaptive sampling are the same sizes as the high-resolution data when they are fed into a super-resolution model, which is a typical procedure in performing super-resolution analysis~\cite{FFT2019a}.
Hence, the data size of the input image is consistent among the dataset, regardless of the grid structure of adaptive-sampled images which varies over each snapshot.
Considering the practical situation of saving (or encoding) these low-resolution images, we suppose using a Run Length Encoding~\cite{RLE}, for example.
Thus, we here consider the number of values that have to be memorized in order to perform RLE for comparing a compression rate of the data, represented as $n_{\rm grid}$.

The reconstruction accuracy with the present adaptive sampling for the $v$ and $w$ components is superior to that of the conventional average pooling while reducing the number of grids.
Especially, the number of grid points for the $w$ component can be saved approximately $75\%$ against that with the conventional pooling.
However, we should be careful in the use of adaptive sampling, as it does not work well for the $u$ component in this particular example.
This is caused by difficulty in determining the universal weights through the convolutional operation for the velocity component whose structures significantly vary over each snapshot.
In turn, there may be an optimal combination of pooling methods and their sampling ratio to achieve the most efficient data saving.

\begin{figure}
	\vspace{0mm}
	\begin{center}
		\includegraphics[width=0.9\textwidth]{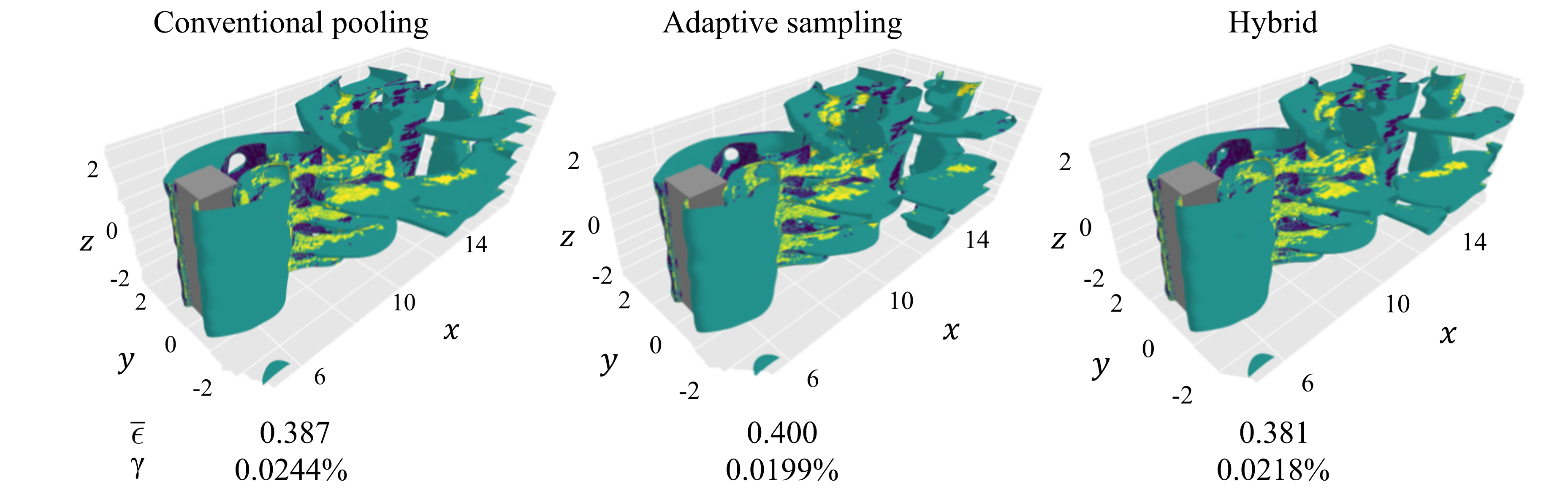}
		\caption{Three-dimensional reconstructed fields from coarse input combining the conventional pooling and the adaptive sampling. Listed values indicate the $L_2$ error, structural similarity and the compression ratio $\gamma$ against the number of grid points over the original three-dimensional discretized domain.}
		\label{fig:3Dfield-adap}
	\end{center}
\end{figure}

Finally, let us exemplify the hybrid method of conventional pooling and adaptive sampling, by presenting the three-dimensional reconstruction in figure~\ref{fig:3Dfield-adap}.
For this investigation, we use $n_{\mathrm{section}}=5$ for all models.
Based on the results discussed above, conventional pooling is used for the $u$ component, while adaptive sampling is used for the $v$ and $w$ components in the present hybrid pooling.
The hybrid utilization can achieve almost the same error level as that with the conventional pooling while saving the storage to approximately 0.0218\% against the original.
This hybrid method can also be combined with optimization to determine the optimal pooling ratio for adaptive sampling, e.g., hyperopt~\cite{BYC2013} and Bayesian optimization~\cite{MMLMBL2019}.

\section{Concluding remarks}
\label{sec:conclusion}

We proposed a method to reconstruct a three-dimensional flow field from its sectional data using a convolutional neural network (CNN). 
The present model was tested with a flow around a square cylinder at ${{Re}}_D=300$.
The proposed 2D-3D CNN model achieved a reasonable reconstruction from a few sections in terms of both the visualization assessments and the qualitative examinations.
We also found that the model trained with as little as $n_{\rm snapshot}=50$ was able to capture large-scale structures of fluid flows.
Considering the application of the present model to experimental situations, the response of the model against noisy input was also investigated, exhibiting promising robustness.

As a further assessment from the aspect of data compression, we combined the three-dimensional reconstruction model with a two-dimensional super-resolution analysis.
Before combining it with the present model, we assessed the super-resolution reconstruction of two-dimensional velocity fields itself. 
The CNN-based super-resolution reconstruction was superior to the bicubic interpolation in terms of estimation accuracy. 
The super-resolved sectional data was then fed into the pre-trained model for three-dimensional reconstruction. 
The case using super-resolved sectional data with downsampling factor $\alpha=8$ can achieve almost the same error level as that using the original two-dimensional velocity data, which corresponds to approximately 1/2000 compression against the original data.

Moreover, we also suggested adaptive sampling which can efficiently downsample data while accounting for the importance of the data to generate a low-resolution field. 
In the present example, the reconstructions for the $v$ and $w$ components with adaptive sampling-based super-resolution analysis were superior to that with the conventional average pooling while being able to reduce the number of grid points, i.e., storage of the data. 
For different problems, however, care should be taken for the choice of an appropriate pooling method depending on the data that users handle.


We finally discuss extensions and the remaining issues concerning our proposal.
One of them is an application to turbulence, although this is extremely challenging as compared to the present demonstration with the square cylinder wake.
To that end, we may be able to capitalize on the idea of autoencoder-based low dimensionalization~\cite{FHNMF2020}, in addition to the present adaptive sampling.
However, we should be careful that there may be a trade-off relationship between the compressibility and explainability of low-dimensionalized representation.
Another limitation is the application to unstructured mesh data.
This limitation is caused by the use of regular CNN in the present study.
Hence, the combination with the state of the arts, e.g., Graph CNN~\cite{ogoke2020graph}, PointNet~\cite{kashefi2020point}, PhyGeoNet~\cite{GSW2020}, and Voronoi CNN~\cite{FukamiVoronoi}, may help us to tackle that issue.
Moreover, in practical applications, how to believe results generated by the machine-learning model is also an important issue since we have no solution data in real-life applications.
In this sense, a preparation for proper uncertainty quantification~\cite{MFRFT2020,MORIMOTO2022133454} needs to be established.
We believe that the present study can serve as one of the pieces toward effective data handling with machine learning for complex nonlinear systems.

\section*{Acknowledgements}

This work was supported by the Japan Society for the Promotion of Science (KAKENHI grant number: 18H03758 and 21H05007). 
We are also grateful to Professor Shinnosuke Obi, Professor Keita Ando, and Professor Linyu Peng, Professor Takuya Kawata for fruitful discussions.
The authors thank Mr. Hikaru Murakami for sharing his DNS code.

\bibliographystyle{unsrt}
\bibliography{references}  
\end{document}